# TUNABLITY OF SOLITARY WAVE PROPERTIES IN ONE DIMENSIONAL STRONGLY NONLINEAR PHONONIC CRYSTALS

C. Daraio[1], V. F. Nesterenko[1,2]*, E. B. Herbold[2], S. Jin[1,2]

*Materials Science and Engineering Program*[1]

*Department of Mechanical and Aerospace Engineering*[2]

University of California at San Diego, La Jolla CA 92093-0418 USA

**Abstract.** One dimensional strongly nonlinear phononic crystals were assembled from chains of PTFE (polytetrafluoroethylene) and stainless steel spheres with gauges installed inside the beads. Trains of strongly nonlinear solitary waves were excited by an impact. A significant modification of the signal shape and an increase of solitary wave speed up to two times (at the same amplitude of dynamic contact force) were achieved through a noncontact magnetically induced precompression of the chains. Data for PTFE based chains are presented for the first time and data for stainless steel based chains were extended into a smaller range by more than one order of magnitude than previously reported. Experimental results were found to be in reasonable agreement with the long wave approximation and with numerical calculations based on Hertz interaction law for discrete chains.

PACS numbers: 05.45.Yv, 46.40.Cd, 43.25.+y, 45.70.-n

# INTRODUCTION

One dimensional strongly nonlinear systems composed by chains of different granular materials are presently a very active area of research because they represent a natural step from weakly nonlinear to strongly nonlinear wave dynamics [1-39]. These systems permit the unique possibility of tuning the wave propagation behavior from linear, to weakly nonlinear and further to strongly nonlinear regimes [19].

In this paper we demonstrate that through simple noncontact magnetically induced precompression it is possible to tune the wave propagation response of the system from the strongly nonlinear to the weakly nonlinear regime. This allows a fine control over the propagating signal shape and speed with an adjustable precompressive force. Novel applications in different areas may arise from understanding the basic physics of these tunable strongly nonlinear 1-D systems, especially at a low amplitude range of stresses corresponding to signals used in ultrasonic diagnostics or in the audible range. For example, tunable sound focusing devices (acoustic lenses), tunable acoustic impedance materials, sound absorption layers and sound scramblers are among the most promising engineering applications [37].

The non-classical, strongly nonlinear wave behavior appears in granular materials if the system is "weakly" compressed with a force $F_0$ [1,2,19]. The term "weakly" is used when the precompression is very small with respect to the wave amplitude. The principal difference between this case and the "strongly" compressed chain (approaching linear wave behavior) is that the ratio of the wave amplitude to the initial precompression is not a small parameter as it was in the latter case. A supersonic solitary wave that propagates with a speed $V_s$ in a "weakly" compressed chain with an amplitude much higher than the initial precompression can be closely approximated by one hump of a periodic solution corresponding to a zero prestress ($\xi_0=0$). This exact solution has a finite length equal to only five particle diameters for a Hertzian type of contact interaction [1,17,19]. In the continuum approximation the speed of this solitary wave $V_s$ has a nonlinear dependence on maximum strain $\xi_m$, which translates to a nonlinear dependence on maximum force $F_m$ between particles in discrete chains. When static precompression ($\xi_0$) is applied the speed of a solitary wave $V_s$ has a nonlinear dependence on normalized maximum strain $\xi_r=\xi_m/\xi_0$ in continuum approximation or on normalized force $f_r=F_m/F_0$ in discrete chain with beads of diameter $a$, bulk density $\rho$, Poisson's ratio $\nu$ and Young's modulus $E$ (Eq. (1), [19,37]):

$$V_s = c_0 \frac{1}{(\xi_r-1)}\left\{\frac{4}{15}\left[3+2\xi_r^{5/2}-5\xi_r\right]\right\}^{1/2} = 0.9314\left(\frac{4E^2 F_0}{a^2\rho^3(1-\nu^2)^2}\right)^{1/6} \frac{1}{(f_r^{2/3}-1)}\left\{\frac{4}{15}\left[3+2f_r^{5/3}-5f_r^{2/3}\right]\right\}^{1/2}. \quad (1)$$

The sound speed $c_0$ in a chain precompressed with a force $F_0$ can be deduced from Eq. (1) at the limit for $f_r=1$:

$$c_0 = \left(\frac{3}{2}\right)^{1/2} c\, \xi_0^{1/4} = 0.9314\left(\frac{2E}{a\rho^{3/2}(1-\nu^2)}\right)^{1/3} F_0^{1/6}, \quad (2)$$

where the constant $c$ is

$$c = \sqrt{\frac{2E}{\pi\rho(1-\nu^2)}}. \qquad (3)$$

Equation (1) also allows the calculation of the speed of weakly nonlinear solitary waves which is the solution of the Korteweg-de Vries equation [19]. When $f_r$ or $\xi_r$ are very large (i.e. $F_0$ ($\xi_0$) is approaching 0) Eq. (1) reduces to the Eq. (4) for solitary wave speed $V_s$ in "sonic vacuum" [3,6] for continuum approximation and discrete chains respectively

$$V_s = \frac{2}{\sqrt{5}} c \; \xi_m^{1/4} = 0.6802 \left( \frac{2E}{a \; \rho^{3/2}(1-\nu^2)} \right)^{1/3} F_m^{1/6}. \qquad (4)$$

For simplicity only the leading approximation was used to connect strains in continuum limit and forces in discrete chain in Eqs. (1) and (4). The similarity between Eq. (2) and Eq. (4) is striking (and can even be misleading) though these equations describe qualitatively different types of wave disturbances. Equation (2) represents the speed of a sound wave with an infinitely small amplitude in comparison with the initial precompression and with a long wave length. This wave is the solution of the classical linear d'Alambert wave equation. Equation (4) corresponds to the speed of a strongly nonlinear solitary wave with the finite width of 5 particle diameters and ratio of solitary wave amplitude to initial precompression equal infinity. This solitary wave is the solution of the strongly nonlinear wave equation first derived in [1]. A strongly nonlinear compression solitary waves exist for any general power law interaction between particles with an exponent in the force dependence on displacement $n>1$ (Hertz law is only a partial case with $n=3/2$) [3,5,10,18,19].

The exponent *n* determines the width of the solitary wave and the dependence of its speed on the maximum strain. The corresponding equations for the speed and width of these solitary waves in continuum approximation were first derived in [3], the results of numerical calculations for discrete chains can be found in [31]. A general type of strongly nonlinear interaction law also supports the strongly nonlinear compression or rarefaction solitary waves depending on elastic "hardening" or "softening" behavior [10,19].

It is possible to find from Eqs. (1)-(4) that the speed $V_s$ of a solitary wave propagating in a one dimensional granular media can be significantly smaller than the sound speed in the material composing the beads and can be considered approximately constant at any narrow interval of its relative amplitude $f_r$ due to the small exponent in the power law dependence. The signal speed of strongly nonlinear and linear compression waves in this condensed "soft" matter can be below the level of sound speed in gases at normal conditions as was experimentally demonstrated for PTFE (polytetrafluoroethylene) based sonic vacuum [37]. The described properties of strongly nonlinear waves may permit the use of these materials as effective delay lines with an exceptionally low speed of the signal.

In stainless steel based phononic crystals, previous experimental and numerical data for the dependence of the solitary wave speed on amplitude exists only at relatively large amplitudes of dynamic forces (20 N – 1200 N) [11,15] and 2 N – 100 N [36] and for the large diameter of stainless steel beads – 8 mm and 26 mm respectively. In the present study we extended the range of experimental data to a lower amplitude range down to 0.1 N to characterize the behavior of stainless steel based strongly nonlinear systems at amplitudes closer to the amplitude of the signals used in ultrasonic diagnostics and for the diameter of beads 4.76 mm. Scaling down of the size of the beads is important for the assembling of practical devices.

**RESULTS AND DISCUSSION**

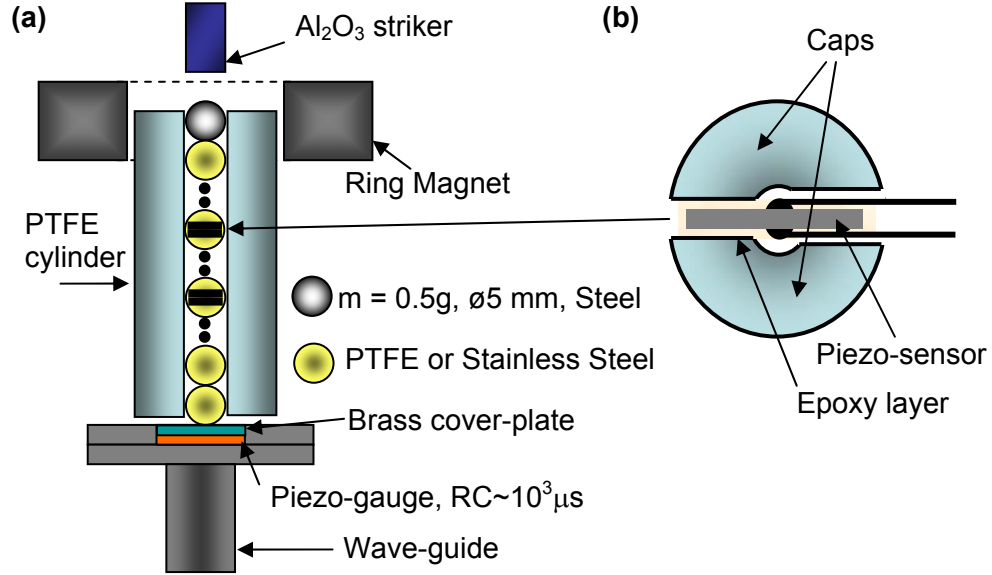

FIG. 1. (a) Experimental set-up for testing of 1-D phononic materials, the magnetic particle on the top is used for magnetic tuning of "sonic vacuum"; (b) Schematic drawing of a particle with embedded piezo-sensor.

Soliton parameters (speed, duration and force amplitude) and reflected pulse from the wall were measured using the experimental set-up presented in Fig. 1. One dimensional PTFE based phononic crystals were assembled in a PTFE cylinder, with inner diameter 5 mm and outer diameter 10 mm, vertically filled with chains of 20 and 52 PTFE balls (McMaster-Carr catalogue) with diameter $a$=4.76 mm and mass 0.1226 g (standard deviation 0.0008 g) (Fig. 1(a)). PTFE has a strong strain rate sensitivity [40] and an exceptionally low elastic modulus in comparison with metals [41,42]. This property can be very attractive to ensure very low speed of soliton propagation and tunability of interfacial properties in laminar composite systems made from chains of different materials. These composite systems exhibit unusual properties with respect to wave reflection [8,19,26] and were proposed as shock protectors [35]. It was demonstrated that chains composed of PTFE beads do support strongly nonlinear solitary waves [37]. Also, chains made from different linear elastic materials like stainless steel, brass, glass [2,11,15,36], and viscoelastic polymeric material like Homalite 100 and nylon [7,9,15] support this type of wave.

For comparison of tunability of signal speed, stainless steel based chains were assembled in the same holder from 20 stainless steel beads (316 steel, McMaster-Carr catalogue) with diameter $a$=4.76 mm and mass 0.4501 g (standard deviation 0.0008 g). A magnetic steel ball, with diameter $a$=5 mm and mass 0.5 g, was then placed on top of the PTFE or stainless steel chains to ensure magnetically induced precompression equal to the weight of the magnet (Fig.1(a)).

Three calibrated piezo-sensors (RC ~$10^3$μs) were connected to a Tektronix oscilloscope to detect force-time curves. Two piezo-gauges made from lead zirconate titanate (square plates with thickness 0.5 mm and 3 mm side) with Nickel plated electrodes and custom microminiature wiring supplied by Piezo Systems, Inc. were embedded inside two of the PTFE and two steel particles in the chains similar to [8,19,36,37]. The wired piezo-elements were glued between two top parts cut from original beads (Fig. 1(b)). This design ensures a calculation of the speed of the signal propagation based on the time interval between maxima detected by different gauges separated by a known distance (usually it was equal to 5 particle diameters) simultaneously with the measurement of the force acting inside the particles. The speed of the pulse was related to the averaged amplitude between the two gauges.

In PTFE, a typical particle with an embedded sensor consisted of two similar caps (total mass $2M=0.093$ g) and a sensor with a mass $m=0.023$ g glued to them. The total mass of these particles was equal to 0.116 g, which is very close to the mass of the original PTFE particle (0.123 g). In theory, the introduction of a foreign element in a chain of particles of equal masses results in weak wave reflections, as observed in [12,18,20]. Numerical calculations with single particle in the chain with a mass 0.116 g embedded into the chain of particles with mass 0.123 g created negligible effects of wave reflections [37]. Therefore such small deviation of the sensor mass from the particle mass (<6%) in both PTFE and stainless steel chains was neglected in the following numerical calculations and chains were considered uniform for simplicity.

The third piezo-gauge, supplied by Kinetic Ceramics, Inc., was bonded with epoxy on two electrode foils for contacts and reinforced by a 1 mm brass plate on the top surface. This sensor was then placed on the top surface of a long vertical steel rod (wave guide) embedded at the other end into a steel block to avoid possible wave reverberation in the system (Fig.1(a)). The sensor embedded in the wall was calibrated using conservation of the linear momentum during a single impact by a steel ball as in [37]. Sensors in the particles were calibrated by comparison with the signal from the sensor at the wall under a controlled, relatively long, simultaneous impact loading.

Considering the particles with an embedded sensor as rigid bodies with all parts moving with the same acceleration, the forces on the sides of the piezoelectric plate working as a sensor can be easily related to the forces acting on the corresponding contacts of the particle with a sensor [37]. This simplification allowed a direct comparison of the force-time curves obtained in experiments with the average of the contact forces derived in the numerical calculations. The forces acting on both sides of the sensor deviate symmetrically from the average value of the contact forces if attenuation is negligible. At the moment when the measured averaged force is maximum $F_{m,e}$ it is equal to the corresponding forces on the contacts of the particle and on both sides of the piezoelement [37].

To relate the maximum value of the force measured by the embedded sensor $F_{m,e}$ and the value of the maximum contact force between neighboring particles $F_m$, which is present in the equations for solitary wave speed (Eq. 1-3), we used the coefficient β:

$$F_m = \beta F_{m,e} + F_0. \qquad (5)$$

The coefficient $\beta$ is determined numerically as $\beta = F_{d,n}/F_{m,n}$. $F_{m,n}$ is the amplitude of the force obtained numerically by averaging, at each time step, the forces on the contacts between adjacent particles. The amplitude of these dynamic contact forces is $F_{d,n}$. $F_{m,n}$ is analogous to $F_{m,e}$ in experiments as described in [37].

The dependence of the coefficient $\beta$ as a function of the normalized force amplitude $F_{m,n}$ with respect to the quasistatic precompression ($F_0$) is shown in Fig. (2). The same coefficient was introduced in [37] for PTFE based systems. The coefficient $\beta$ is independent of the elastic modulus, but varies with respect to the amplitude of the solitary wave. This dependence was taken into account when calculating the value of dynamic contact force from measured values by gauges inside particles for both PTFE and stainless steel cases.

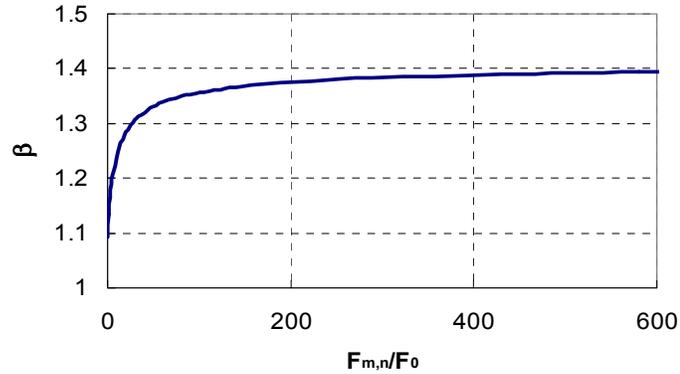

FIG. 2. Plot representing the $\beta$-coefficient found in numerical calculations for the stainless steel case, as a function of the normalized maximum averaged dynamic force on the contacts ($F_{m,n}$ is corresponding to $F_{m,e}$ measured by embedded gauges in experiments).

Strongly nonlinear sonic vacuum type materials allow significant change of speed of propagation with initial prestress which make them qualitatively different from linear systems. In the present case the tunability of the signal propagating through the chains was achieved by adding a magnetically induced preload. In contrast, previous investigations used a preloading of the system through the mechanical application of the load using a wire with attached weights or contact interaction between the last particle and the frame [4,11]. The former type of preloading allows application of a constant external force to the end magnetic particle independently from its displacement. This was difficult to achieve in the previous attempts especially if the dynamic displacement of the end particle was comparable to its initial displacement due to the preloading. This case would change boundary conditions in an uncontrollable way. The proposed magnetic precompression ensures fixed boundary conditions which are used in numerical calculations and even allows the additional possibility of a fine tuning of the preload which can be time dependent. A Neodymium-Iron-Boron ring magnet (internal and external diameters 11.43 mm, 53.34 mm respectively, with mass 242.2 g from Dexter Magnetics Technologies) was placed around the PTFE cylinder containing the chain and held in place by the magnetic interaction with the magnetic steel particle placed on the top of the chain (Fig. 1(a)). Throughout the whole chain the precompressive force so obtained was 2.38 N. This particular set up was chosen to allow an easy noncontact control over the variation of the preload by slight axial movements of the magnetic ring or by the use of ring magnets of different masses. A higher level of magnetically induced precompression in experiments (4.25 N) was achieved with added mass to the magnet.

In the absence of the magnetic precompressive loading a non-uniform gravitational prestress of the chains caused by the weight of the particles composing the system was taken into account.

Nonmagnetic alumina ($Al_2O_3$) cylinders (0.47 g and 1.2 g) were used in experiments to generate pulses of different durations and amplitudes in the 1-D phononic crystals, by controlled impacts on the top particle of the chain. Waves of different amplitudes were excited with different velocities of the striker impacts.

For comparison with experiments numerical calculations were performed using Matlab. Particles were treated as rigid bodies connected by nonlinear springs according to Hertz law as in previous investigations [1,2,4,14,19]. As mentioned earlier, to compare numerical results and experimental data, we calculated an average compression force for solitary waves based on the corresponding compression forces on the particle contacts.

For the PTFE beads, Poisson ratio was taken equal to 0.46, the density 2200 Kg/m$^3$ and the value of the elastic modulus was taken equal to 1.46 GPa based on published data and our previous experimental results [37,43]. Poisson's coefficient for stainless steel 316 was taken equal to 0.3, density 8000 kg/m$^3$ and the elastic modulus equal to 193 GPa [44]. Hertz law was also used to describe the interaction between the flat wall at the bottom of the chain and last particle. In this case, the elastic modulus for red brass (Cu85-Zn15) was equal 115 GPa and Poisson's ratio at 0.307 [45]. Similarly, for the alumina impactor we had $E$=416 GPa, $v$=0.231 [46]. The nonuniform precompression resulting from the gravitational loading of the particles in the vertical chains was included in calculations and added to the magnetically induced uniform prestress. Energy and linear momentum (before interaction with a wall) were conserved with a relative error of 10$^{-8}$ % and 10$^{-12}$ % similarly as in [37]. Dissipation was not included in the current numerical analysis.

The amplitude of the solitary waves was measured with accuracy in the range of 15% to 30%. The larger error at lower amplitudes is attributed to the higher signal to noise ratio.

The accuracy of the speed measurements can be estimated within 10 % due to the uncertainty in the sensors alignment (about 1 mm for each sensor).

It is well known that in "sonic vacuum" type systems trains of solitary waves are rapidly produced due to the decomposition of an initial pulse on a distances comparable to the solitary wave width [1,2,4,19]. Strong dissipation in the system may obscure the appearance of this phenomenon or even completely suppress it. The moderate dissipation allowed to demonstrate this property experimentally for steel based phononic crystals [2,19] and recently also for PTFE based systems [37]. In PTFE based sonic vacuum, dissipation was more noticeable than in the steel beads chain especially in the case of small diameter particles (2.38 mm diameter) where the splitting of the initial pulse into a train of solitary waves was less pronounced than in the numerical calculations (Figs. 3(c) and (d) in [37]). The number of solitary waves with significant amplitude can be estimated as the ratio of impactor mass to the mass of the bead in the chain [1,19,39]. In the present case the mass of the alumina striker was chosen to be about four times the mass of the particles in the chain, producing a train of four solitary waves. It should be mentioned that the number of solitary waves may be significantly larger if smaller amplitudes are also included as found in numerical calculations [39].

The typical results of the experiments for a chain of 52 PTFE beads capped by the magnetic steel particle are shown in Fig. 3 (a), (c). In these experiments, the impulse loading was generated using a nonmagnetic 0.47g $Al_2O_3$ cylinder with a velocity of 0.89 m/s impacting on the top magnetic particle. Internal sensors were placed in the $20^{th}$ and $29^{th}$ beads from the top magnetic particle and one more piezo-gauge was embedded in the wall. The distance between the two internal sensors was equal 43.1 mm. The zero time in experiments corresponding to Fig. 3 (a),(c) is arbitrary chosen (the start of recording triggered by the signal), while the zero time in numerical calculations, Fig. 3 (b),(d) corresponds to the moment of impact. The numerical calculations took into account the presence of the magnetic bead at the top of the chain in all cases.

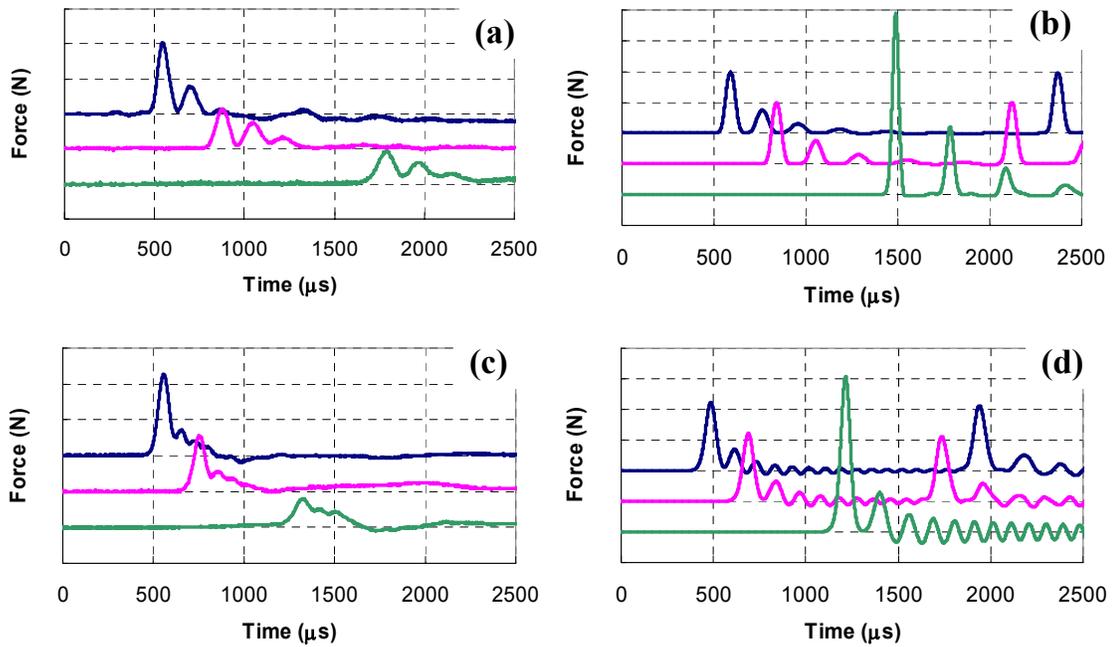

FIG. 3. Experimental and numerical results corresponding to a chain of PTFE beads with and without magnetic precompression. (a) Forces detected in experiment by the sensor in the 20$^{th}$ ball from the impacted side (top curve), by the sensor in the 29$^{th}$ ball (middle curve) and at the wall (bottom curve) without magnetic pre-compression, vertical scale is 0.5 N; (b) Numerical calculations of averaged dynamic contact forces between particles and at the wall corresponding to the signals detected by the sensors in (a), including gravitational precompression, vertical scale is 2 N; (c) Forces detected in experiment with magnetic pre-compression equal 2.38 N, vertical scale is 1 N; (d) Numerical calculations corresponding to experimental conditions in (c) including gravitational and magnetic precompression, vertical scale is 2 N.

The obtained experimental profiles and corresponding numerical results for gravitationally loaded chain are presented in Fig. 3(a),(b). In this weakly compressed state, the Teflon-based "sonic vacuum" demonstrates the typical fast decomposition of the initial impulse on a distance comparable with the soliton width similar to the trains observed in a shorter 21 particle chain (Fig. 5 in [37]). In the Fig. 3(a) the duration of the leading solitary wave pulse was equal to 170 μs. The speed of its propagation, based on the time interval between the signal's maxima measured by the two embedded sensors, was 129 m/s. These values result in a width of the solitary wave equal to 4.6 particles diameters. This is in a reasonably good agreement with the solitary wave width predicted in long wave theory for a chain of particles interacting according to Hertz law (5 particles diameter) [1,19]. These parameters allow one to evaluate the characteristic strain rates for PTFE deformation at the contact area equal to $4 \times 10^2$ s$^{-1}$ [37]. A reasonably good agreement of the experimental results (Fig. 3(a), (c)) with the numerical calculations at the same testing conditions (Fig. 3(b),(d)) was found for this and other investigated conditions of impacts. Although, the important role of dissipation is demonstrated especially when comparing numerical and experimental results for the force on the bottom of the chain (Fig. 3).

Results of numerical calculations demonstrated that magnetic precompression under similar type of impact caused a qualitative change of the signal shape from a train of solitary waves in a weakly compressed case (Fig. 3(b)) to a few solitary waves with a periodically oscillating tail (Fig. 3(d)). In experiments (Fig. 3(c)) the periodic oscillatory tail was not observed, which may be due to the interaction between the wall of the holder and the particles.

One of the most interesting consequences of the noncontact magnetic prestress acting on the chain is a significant change in the speed of the signal propagation. In Fig. 3(a-d) the pulse excited by the same impact resulted in an increased speed from a value of 129 m/s for the uncompressed chain to 222 m/s for the magnetically loaded system. This translates in a ~70% control over the initial "natural" signal speed of the strongly nonlinear system. This speed of solitary wave in uncompressed PTFE chains is about two times smaller than speed of solitary wave in the chain of nylon beads [15] with amplitude 1 N due to a smaller elastic modulus of Teflon in comparison with Nylon. This is in agreement with theoretical dependence of speed on elastic modulus.

Interestingly, in numerical calculations the dynamic amplitudes of the incident pulses in the magnetically prestressed chains are close to the one observed in the only gravitationally loaded chains (compare corresponding curves in Fig. 3(b) and (d)). Meanwhile, the dynamic amplitude on the wall in the magnetically precompressed chains in numerical calculations is noticeably smaller than the one in the former case (see bottom curves in Fig. 3(b) and (d)). This shows that the precompression can significantly reduce the maximum of the dynamic force acting on the wall though this phenomenon was not observed in the experiments probably due to the decrease of dissipation in the prestressed chain.

The same impact in experiments resulted in an increased amplitude of the incident pulses propagating in the preloaded PTFE chain by 2 times (compare corresponding curves in Fig. 3(a) and (c)). This is probably also due to the significant decrease of dissipation related to the increase of the stiffness and to the reduction of the relative displacement between the particles under similar conditions of loading at higher precompression.

The low amplitude precursor under magnetic precompression was noticeable especially in the wall gauge (Fig. 3(c), bottom curve). This may be due to the propagation in front of the leading solitary wave of a high speed disturbance with the speed of sound in solid PTFE through the central cylindrical column created when the initial precompression flattens the contact area between spherical particles. This phenomenon is not included in the used numerical approach and requires further investigation.

Table 1 shows a summary of the typical results for the amplitude of the dynamic contact forces $F_d$ and corresponding velocities $V_s$ obtained in experiments. A comparison of 5 representative values obtained experimentally with the corresponding values obtained in numerical calculation and with the theory of long wave approximation for given values of $F_d$ are included for clarity.

Table 1. Comparison of experimental values of the solitary wave speeds $V_s$ with the corresponding values obtained in numerical calculation and with the theoretical values from long wave approximation.

| Only Gravitational Precompression | | | | | | Magnetic and Gravitational Precompression | | | | | |
|---|---|---|---|---|---|---|---|---|---|---|---|
| Experimental Data | | Numerical Results | | Long Wave Approximation | | Experimental Data | | Numerical Results | | Long Wave Approximation | |
| $F_d$ [N] | $V_s$ [m/s] | $F_{d,n}$ [N] | $V_s$ [m/s] | $F_d$ [N] | $V_s$ [m/s] | $F_d$ [N] | $V_s$ [m/s] | $F_{d,n}$ [N] | $V_s$ [m/s] | $F_d$ [N] | $V_s$ [m/s] |
| 5.5 | 168 | 5.7 | 184 | 5.7 | 182 | 5.7 | 241 | 5.8 | 228 | 5.8 | 230 |
| 4.3 | 166 | 4.3 | 177 | 4.3 | 174 | 4.5 | 235 | 4.5 | 224 | 4.5 | 227 |
| 3.3 | 167 | 3.3 | 170 | 3.3 | 167 | 3.4 | 230 | 3.2 | 220 | 3.2 | 223 |
| 2.5 | 163 | 2.6 | 163 | 2.6 | 161 | 2.6 | 221 | 2.6 | 218 | 2.5 | 221 |
| 2.1 | 143 | 2.0 | 159 | 2.0 | 155 | 2.0 | 223 | 2.0 | 215 | 2.0 | 220 |

The force data presented in the experimental part of the table ($F_d$) represents the average of the two contact force amplitudes $F_{d,e}$ extracted from the amplitudes of the corresponding signals of the two embedded gauges with amplitudes $F_{m,e}$, $F_{d,e} = \beta F_{m,e}$. The solitary wave speeds related to $F_d$ were obtained dividing the distance between the sensors (about 5 particle diameters) by the measured peak-to-peak time interval. To get the speed of a solitary wave corresponding to the given contact force amplitude $F_d$ in the long wave approximation we used Eq. (1) where total force $F_m$ is the sum of $F_d$ plus $F_0$ caused by initial precompression (gravitational or gravitational plus magnetic).

It should be noted that the speeds of the solitary wave at the investigated dynamic amplitude (which is about 2 times higher than initial precompression force) in the magnetically precompressed chain is slightly higher than the speed of the sound in the system $c_0 = 211.2$ m/s evaluated from Eq. (3).

The comparison of the shapes of the leading pulses detected experimentally and the stationary solitary waves obtained in numerical calculations is shown in Fig. 4.

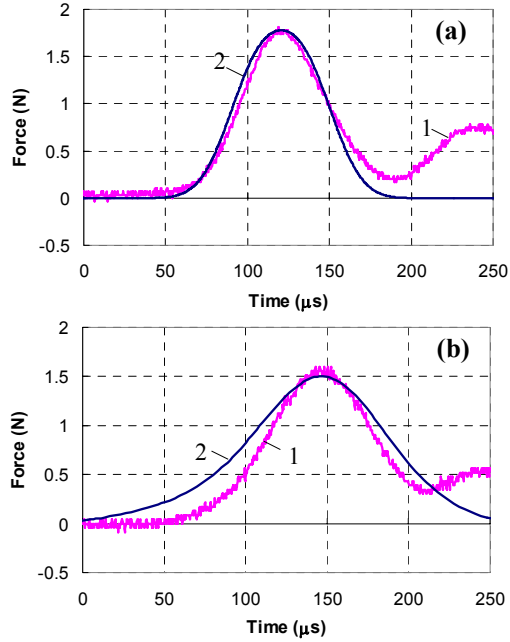

FIG. 4. The leading pulses obtained in experiments (curves (1)) and stationary solitary waves found in numerical calculations (curves (2)) with the same amplitude. (a) PTFE chain under gravitational precompression and the stationary solitary wave calculated in a "sonic vacuum", (b) Magnetically precompressed PTFE chain and the stationary solitary wave calculated in a uniformly precompressed chain with force 2.38 N. In both cases the PTFE chain was composed of 52 particles and sensor was embedded into the 21-st particle

from the top, a magnetic steel particle was positioned on the top and the impactor was the 0.47 g $Al_2O_3$ cylinder.

From Fig. 4 we can see that the shape of the leading pulses in experiments is close to the shape of the stationary solitary waves in the numerical calculations despite that in experiments these pulses were not completely separated from the oscillatory tail at the given distances of the sensors from the impacted end of the chain.

It should be mentioned that in the precompressed chains the splitting of the incident pulse into a train of solitary waves is delayed especially when the dynamic amplitude of the pulse is smaller or comparable with the initial precompression [4,19,27]. At relatively small value of their ratio the strongly nonlinear solitary wave become close to the soliton solution of the Korteweg-de Vries wave equation with width infinitely increasing with amplitude decreasing to zero [4,19]. Detailed analysis of the depth for solitary wave stabilization depending on the relative amplitude of the solitary wave and initial precompression can be found in [27]. In our numerical calculations the leading solitary pulse with an amplitude of 1.5 N, therefore smaller than the magnetic precompression force $F_0$=2.38 N (curve (2) in Fig. 4(b)), was completely separated from the oscillatory tail in the region about the 950$^{th}$ particle from the impacted end (curve 2 in Fig. 4(b) corresponds to the 980$^{th}$ particle in the chain). This may explain why the stationary solitary pulse in numerical calculations is wider than the width of the leading pulse in magnetically precompressed chain experimentally detected at the 20$^{th}$ particle (Fig. 4(b)). The corresponding increase of the width of the leading pulse when it propagates inside the magnetically precompressed chain was observed in numerical calculations.

In the strongly nonlinear regime of pulse propagation in a gravitationally precompressed chain the separation was already completed at the 20$^{th}$ particle (Fig. 3(b)). An increase of the amplitude of the pulse to the level of 5 N resulted in numerical calculations in a faster splitting into a train of solitary waves at a distance about 20 particles in a magnetically precompressed chain (Fig. 3(d)). The presence of dissipation in experiments delays the process of pulse disintegration into the train of solitary waves in comparison with numerical calculations (compare Figs. 3(a), (b) and Figs. 3(c), (d)). The dissipation was increased at larger amplitudes of the pulses.

The speed of a solitary wave in the long wave approximation is slightly lower than in the numerical calculations for a gravitationally loaded discrete chain at the same amplitude of dynamic force $F_d$ (see left side of Table 1). The opposite tendency is characteristic for the case where additional magnetically induced loading is applied (see right side of Table 1). In both cases the difference between the long wave approximation and the results for a discrete chain is actually very small in agreement with the results of [31]. The experimental results agree with the data from the numerical calculations and the long wave approximation.

Figure 5 shows the comparison of the experimental results with the theoretical values obtained from the long wave approximation, and the numerical data for discrete chains of both gravitationally (lower curves and experimental points) and magnetically precompressed (2.38 N) systems (upper curves and corresponding experimental points). It is known that the theoretically predicted speed of solitary waves in strongly nonlinear phononic crystals has a strong dependence on the amplitude represented by Eqs. (1)-(3) for "weakly" precompressed chains. The curves based on the long wave approximation (curves 2 and 4) are very close to the one obtained in numerical calculations (curves 3 and 5). It should be mentioned that this good agreement is obtained when only the leading approximation was used to connect strains in continuum limit and forces in discrete chain in Eqs. (1) and (4). A reasonably good agreement between the theoretical predications and the experiments is observed. Experimental data obtained with increased precompression (4.25 N, not shown in Fig. 5) also followed the general trend prescribed by long wave approximation.

At higher amplitudes of the solitary wave the agreement between the theoretical data and experiments is better than at lower amplitude for the weakly (gravitationally) precompressed chain. We attribute this behavior to the dependence of the elastic modulus of the dynamically deformed PTFE on strain (strain rate) on the particle contact [37].

It should be mentioned that for PTFE particles with a dynamic elastic modulus 1.46 GPa the dependence of the solitary wave speed on the dynamic amplitude $F_d$ is actually very close to the long wave approximation for sonic vacuum (Eq. (4)), if the initial gravitational preload is neglected and the dynamic force is associated with the force $F_m$.

Gravitational precompression though being very small in the middle between two gauges (about 0.017 N) results for the long wave approximation in a sound speed in the chain equal to 92.6 m/s with a dynamic elastic modulus for PTFE taken equal to 1.46 GPa. This value determines the theoretical limit of the solitary wave in the case of $F_d$ approaching zero for a gravitationally precompressed chain in the long wave approximation. The numerical data also demonstrated the same limit. In the low amplitude range the experimental data is below the long wave approximation and numerical curves. This may be due to the lower than 1.46 GPa dynamic elastic modulus of PTFE at smaller strains (strain rates) [37].

The increase of the sound speed $c_0$ in the PTFE chain under the added magnetically induced precompression (2.38 N) in comparison with only gravitational preload (0.017 N, $c_0$=92.6 m/s) is about 128% at practically the same density. This results in a corresponding change of acoustic impedances.

It should be pointed out that the curve 1 for sonic vacuum in Fig. 5 has a limit equal to zero at zero $F_d$, which is not shown in the Fig. 5 due to the steep decrease of the solitary wave speed with amplitude in this case.

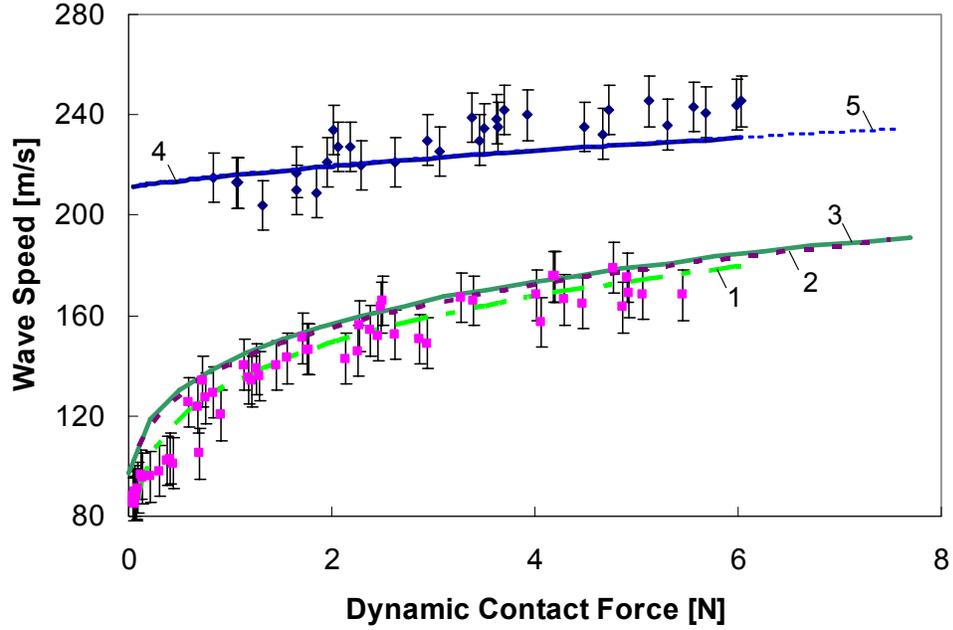

FIG. 5. Dependence of the solitary wave speed on the amplitude of the contact dynamic force for gravitationally loaded and for magnetically tuned chains composed of PTFE beads. The experimental values for corresponding curves are shown by solid squares and dots. Curve 1 represents the long wave approximation for the "sonic vacuum" at elastic modulus equal 1.46 GPa. Curves 2 and 4 are the theoretical curves based on Eq. 1, and the corresponding numerical calculations for discrete chains are represented by curves 3 and 5.

The agreement between the speeds in the magnetically tuned system and the estimated theoretical values means that the equations for the dependence of the solitary wave speed based on the nonlinear continuum approximation can be a reasonable foundation for the design of "sonic vacuum" type devices based on PTFE beads.  The speed of the solitary wave at the minimum investigated dynamic amplitude in the magnetically precompressed chain (limit at $F_d$ approaching zero for curves 4 and 5 for long wave approximation and numerical calculations correspondingly) is very close to the speed of the sound in the system $c_0 = 211.2$ m/s estimated based on long wave approximation.

In these experiments the maximum total force (8.4 N) was larger than in the gravitationally precompressed chain.  Additionally PTFE was loaded differently in comparison with the weakly compressed chain even when the maximum forces were the same. As mentioned, the elastic modulus of PTFE chains was selected to have a constant value of 1.46 GPa from the extrapolation of the Hugoniot data [43].  Despite a difference in the loading conditions at this large strain the higher value of the elastic modulus (compared to the ultrasonically measured elastic modulus at normal conditions) is able to describe the experimental data. Further research on the value of elastic modulus of PTFE under conditions of dynamic deformation and application of Hertz approximation for the contact law is necessary to clarify the observed behavior.  The dependence of PTFE's elastic modulus on contact strain may cause deviation of the contact law from Hertzian type behavior and deviation from a power law with an exponent equal to 3/2.

To compare the predictions of the long wave theory of the tunability of strongly nonlinear systems with the variation of the elastic modulus of the beads, we measured the parameters of the solitary waves in a stainless steel based system under the same magnetically induced precompression.  Stainless steel particles have more than two order of magnitude difference in the elastic modulus (193 GPa) in respect to PTFE (1.46 GPa). The experimental results of previous research [2,11,15,36] are in excellent agreement with the predictions of the long wave theory and with the results of a numerical calculations for chains composed of stainless steel beads. The data presented in this paper explores a different noncontact method for the application of an external preload on the chain of particles with smaller diameters. Magnetically induced preload (Fig. 1(a)) is essential to ensure the application of a constant external force independent from the motion of the end particle. In addition the diapason of the solitary wave amplitudes was shifted toward the low range where almost no systematic experimental data are available for preloaded chains or for chains with this relatively small diameter of particles. This is very important for the design of tunable devices based on the concept of a "sonic vacuum".

Experimental and numerical results for chains composed of stainless steel beads with diameter 4.76 mm are shown in Fig. 6. Impacts were generated dropping an $Al_2O_3$ cylinder of 1.2 g on the top magnetic steel particle (0.5 g). Solitary waves in the chain of stainless steel beads also demonstrated a significant increase of speed with added magnetically induced prestress of 2.38 N. For example, in experiments the leading solitary pulse presented in Fig. 6(a) had a speed of 580 m/s in the only gravitationally precompressed chain which changed to 688 m/s in the magnetically precompressed one Fig. 6(c). A similar behavior was observed in the numerical calculations (Fig. 6(b) and (d)).

Similarly to the PTFE based chains, the numerical calculations in the stainless steel based system did not show any change of the dynamic force amplitudes of the incident solitary pulses with the application of the initial magnetically induced precompression (see Fig. 6(b), (d)). This behavior is different than the one observed in experiments, where the added precompression did change the amplitude of the incident pulse (compare Fig. 6(a) and (c)). This difference can be explained in experiments by the reduction of dissipation in the precompressed chain. This reduction may be due to the decrease of the relative displacement of the beads during the wave propagation. In particular, this is evident when comparing the experimental and numerical results of the precompressed chains. Here the amplitudes of the incident pulses are practically the same (compare Fig. 6(c) and (d)). In contrast, the maxima of the forces of the reflected waves are significantly smaller in the experiments than in numerical calculations for a precompressed chain. This reduction of dissipation in the precompressed system was also observed in the PTFE chains, although here the overall dissipation was larger because of the viscoelastic nature of PTFE in comparison with the elastic behavior of stainless steel.

It is interesting that a tendency toward a delay of the splitting into a solitary wave train was observed when the amplitude of the incident pulse was increased which may be due to the dissipation increase with amplitude of the pulse.

A slight change of the propagating pulse shape was observed in the magnetically precompressed chain. The reduced time interval between the maxima of the incident solitary waves in experiments and in numerical calculations (Figs. 6(c) and (d) respectively) demonstrates a significant delay of the pulse splitting. This effect in stainless steel based systems is less noticeable than in the PTFE chains (see Fig. 3(c) and (d)) which may be due to the smaller speed of signal propagation in the PTFE chains. Also, in the stainless steel case, no increase of the wall's pulse amplitude was detected, contrarily to the PTFE based chains. The larger difference between the elastic constant of the wall and the one of the PTFE chain can be the cause of this phenomenon.

The addition of a prestress in chains composed of steel beads was also examined in [11] but for a range of force amplitudes much higher than the one presented in this study. This paper also extends the investigation to multiple solitary waves in contrast to single solitary wave type pulses in [11].

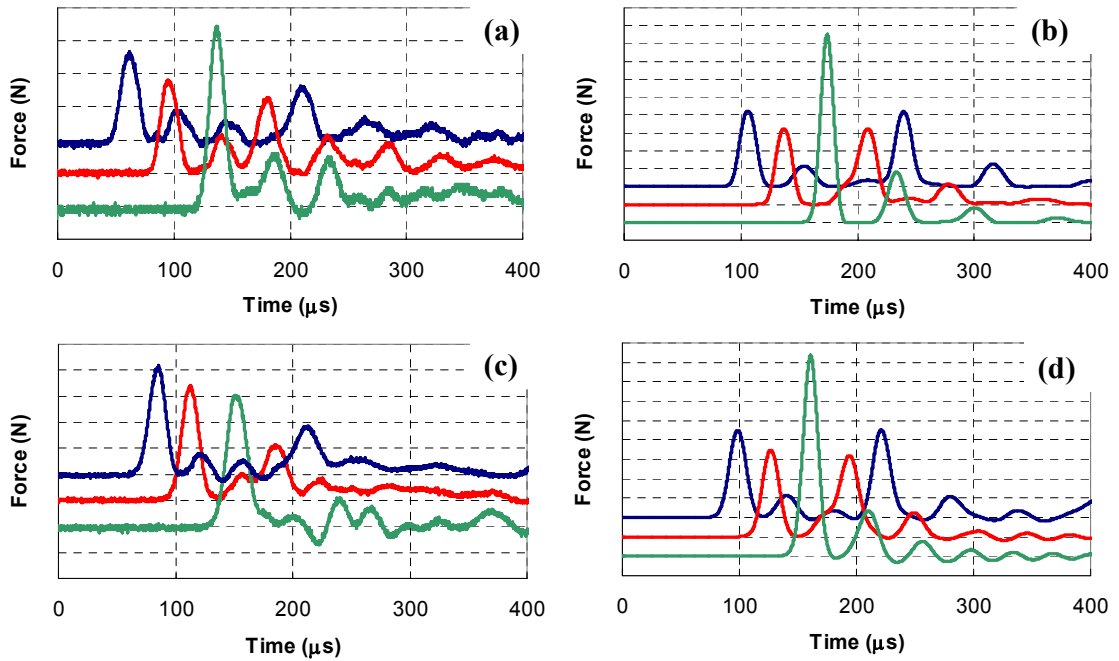

FIG. 6. Experimental and numerical results on a chain of 21 stainless steel (316) beads with and without magnetic precompression impacted by alumina striker with a mass

equal 1.2 g and velocity 0.44 m/s. (a) Force detected in experiment by the sensor in the $9^{th}$ ball from the wall (top curve), by the sensor in the $5^{th}$ ball (middle curve) and at the wall (bottom curve) without magnetic pre-compression; (b) Numerical calculations corresponding to experimental conditions in (a), including gravitational precompression; (c) Force detected in experiment with magnetic pre-compression equal 2.38 N; (d) Numerical calculations corresponding to experimental conditions in (c), including gravitational and magnetic precompression. Vertical scale is 5 N for all figures.

Figure 7 shows the comparison of the experimental results with the theoretical values for solitary wave speed versus dynamic force amplitude obtained from the long wave approximation, and the numerical data for discrete chains of both gravitationally (lower curves and experimental points) and magnetically precompressed systems (upper curves and corresponding experimental points). The curves corresponding to the stainless steel system extend to a wider range of amplitudes and velocity and the scale is therefore wider than in Fig. 5 for PTFE chains. In this plot, the curves for the long wave approximation and the numerical results coincide for both gravitationally (curve 1 in Fig. 7) and magnetically precompressed chains (curve 2 in Fig. 7). The pulse speed at dynamic amplitude approaching zero in the magnetically and gravitationally precompressed chain (2.44 N) is approaching value of sound speed ($c_0$= 539.6 m/s) derived from long wave approximation (Eq. 2). The change of the sound speed $c_0$ in the stainless steel chain under the added magnetically induced precompression (2.38 N) in comparison with the system under only the gravitational preload (0.062N, $c_0$=292.6 m/s) is about 84% at practically the same density, resulting in the corresponding change of acoustic impedances.

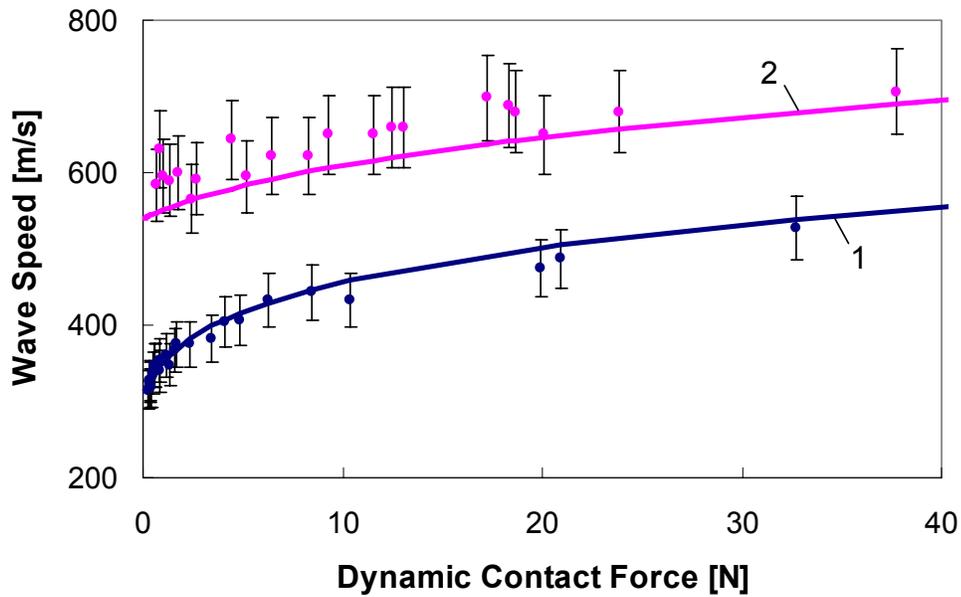

FIG. 7. Dependence of the velocity of solitary wave on amplitude of dynamic force for gravitationally loaded and for magnetically tuned stainless steel chain. Experimental values for corresponding curves are shown by solid dots. Curves 1 and 2 are the theoretical curves based on Equation (3), corresponding numerical calculations for discrete chain are practically coinciding with curves 1 and 2 based on long wave approximation.

These results show that stainless steel particles under the same precompression force demonstrate a larger absolute increase of the solitary wave speed in comparison to the PTFE system (compare Fig. 5 and 7) both in experiments, theory and in numerical calculations. Because the PTFE system is elastically much softer, this result can appear counterintuitive. This behavior of the solitary wave speed with preloading is due to the fact that, in general, a smaller displacement under the same precompression is overweighed by a larger elastic modulus of stainless steel.

These properties of strongly nonlinear phononic crystal can be used for controlled impulse transformation in relatively short transmission lines as well as in tunable acoustic focusing lenses.

## CONCLUSIONS

A new method of preloading phononic crystals via magnetic interaction was successfully demonstrated for one dimensional strongly nonlinear systems based on two qualitatively different materials – elastic (stainless steel) and viscoelastic (PTFE) beads having more than two orders of magnitude difference in elastic moduli. This novel method of precompression ensured well controlled boundary conditions and the possibility of a time-dependent fine tuning of the signal speed. The change of the sound speed in these systems under the investigated magnetically induced precompression in comparison with the gravitational preload only is 84% and 128% for stainless steel and PTFE chains correspondingly at practically the same density. They result in the corresponding change of acoustic impedances.

Both systems were investigated for different conditions of loading under the tunable unobtrusive magnetic prestress. A significant tunability of the speed of the signal was achieved over the "natural" uncompressed strongly nonlinear system. The change of the solitary wave speed with prestress in experiments is very well described by the results of the long wave approximation and the numerical data.

The initial preloading significantly reduced dissipation in the experiments in both PTFE and stainless steel based systems. A decrease of the reflected wave amplitude on the wall was observed in numerical calculations for PTFE system. A delay of the splitting of the solitary waves under prestress was observed in experiments and in numerical calculations.

The solitary wave properties obtained in the long wave approximation are a reliable tool in the designing of sonic vacuum based devices.

## ACKNOWLEDGEMENTS

This work was supported by NSF (Grant No. DCMS03013220).